\documentclass[12pt,onecolumn]{aastex62}
\usepackage{amsmath}
\usepackage{graphicx}
\usepackage{multirow}
\usepackage{color}
\usepackage{natbib}
\usepackage{morefloats}
\begin{document}

\title{The Effects of Filter Choice on Outer Solar System Science with LSST}

\author[0000-0001-8736-236X]{Kathryn Volk}
\correspondingauthor{Kathryn Volk}
\email{kvolk@lpl.arizona.edu}
\affiliation{Lunar and Planetary Laboratory, The University of Arizona, Tucson, USA}

\author[0000-0003-4365-1455]{Megan E. Schwamb}
\affiliation{Gemini Observatory, Northern Operations Center, Hilo, HI USA}

\author[0000-0001-6680-6558]{Wesley C. Fraser} 
\affiliation{Astrophysics Research Centre, Queen's University Belfast, Belfast, United Kingdom}

\author[0000-0002-6702-7676]{Michael S.~P.~Kelley}
\affiliation{University of Maryland at College Park, College Park, MD, USA}

\author[0000-0001-7737-6784]{Hsing~Wen~(Edward)~Lin}
\affiliation{Department of Physics, University of Michigan, Ann Arbor, MI, USA}

\author[0000-0003-1080-9770]{Darin Ragozzine}
\affiliation{Brigham Young University, Provo, UT, USA}
 
 \author[0000-0001-5916-0031]{R. Lynne Jones}
\affiliation{University of Washington, Seattle, WA, USA}

 \author[0000-0001-9328-2905]{Colin Snodgrass} 
 \affiliation{University of Edinburgh, Edinburgh, UK}

\author[0000-0003-3257-4490]{Michele T. Bannister}
\affiliation{Astrophysics Research Centre, Queen's University Belfast, Belfast, United Kingdom}

\date{November 30, 2018}

\begin{abstract}
        Making an inventory of the Solar System is one of the four pillars that the requirements for the Large Synoptic Survey Telescope (LSST) are built upon. 
        The choice between same-filter nightly pairs or different-filter nightly pairs in the Wide-Fast-Deep (WFD) Survey will have a dramatic effect on the ability of the Moving Object Pipeline System (MOPS) to detect certain classes of Solar System objects; many of the possible filter pairings would result in significant ($\sim50\%$ or more) loss of Solar System object detections.
        In particular, outer Solar System populations can be significantly redder than those in the inner Solar System, and nightly pairs in $r$-band will result in the deepest survey for the outer Solar System. 
        To maximize the potential for outer Solar System science, we thus advocate for ensuring that the WFD survey contains a sufficient number of $r$-$r$ nightly pairs for each field during a discovery season to ensure detection and linking using MOPS.  
        We also advocate for adding additional spectral energy distributions (SEDs) that more accurately model outer Solar System populations to the pipeline for evaluating the outputs of the LSST operations simulator.
        This will enable a better estimate of how many Solar System population detections are lost or gained for different filter choices in the WFD survey.
        
\end{abstract}

\section{White Paper Information}

\begin{enumerate} 
\item {\bf Science Category:} Taking an Inventory of the Solar System
\item {\bf Survey Type Category:} the main `wide-fast-deep' survey
\item {\bf Observing Strategy Category:} an integrated program with science that hinges on the combination of pointing and detailed observing strategy
\end{enumerate}

\clearpage
\section{Scientific Motivation}\label{sec:motivation}

One of the main motivations of the Large Synoptic Survey Telescope (LSST) is to provide an inventory of the Solar System \citep{2008arXiv0805.2366I,2009arXiv0912.0201L}. 
To achieve the goals in the LSST Solar System Science Collaboration's roadmap \citep{2018arXiv180201783S}, LSST must discover large numbers of small bodies in the Solar System.
Solar system object detection will be performed with the LSST project's Moving Object Processing System (MOPS) (see details in LSE-30\footnote{\url{http://ls.st/LSE-30}} and  LDM-156\footnote{\url{http://ls.st/LDM-156}}).
MOPS requires at least two visits per night in a field to detect and identify moving objects.
To make new discoveries and link discovery of known Solar System bodies to prior observations at 95\% confidence, MOPS requires three tracklets (a pair of images in the same night, acquired no more than 90 minutes apart) acquired within a 15 day span. 

\begin{figure}[h!]
\begin{center}

	\includegraphics[width=0.6\columnwidth]{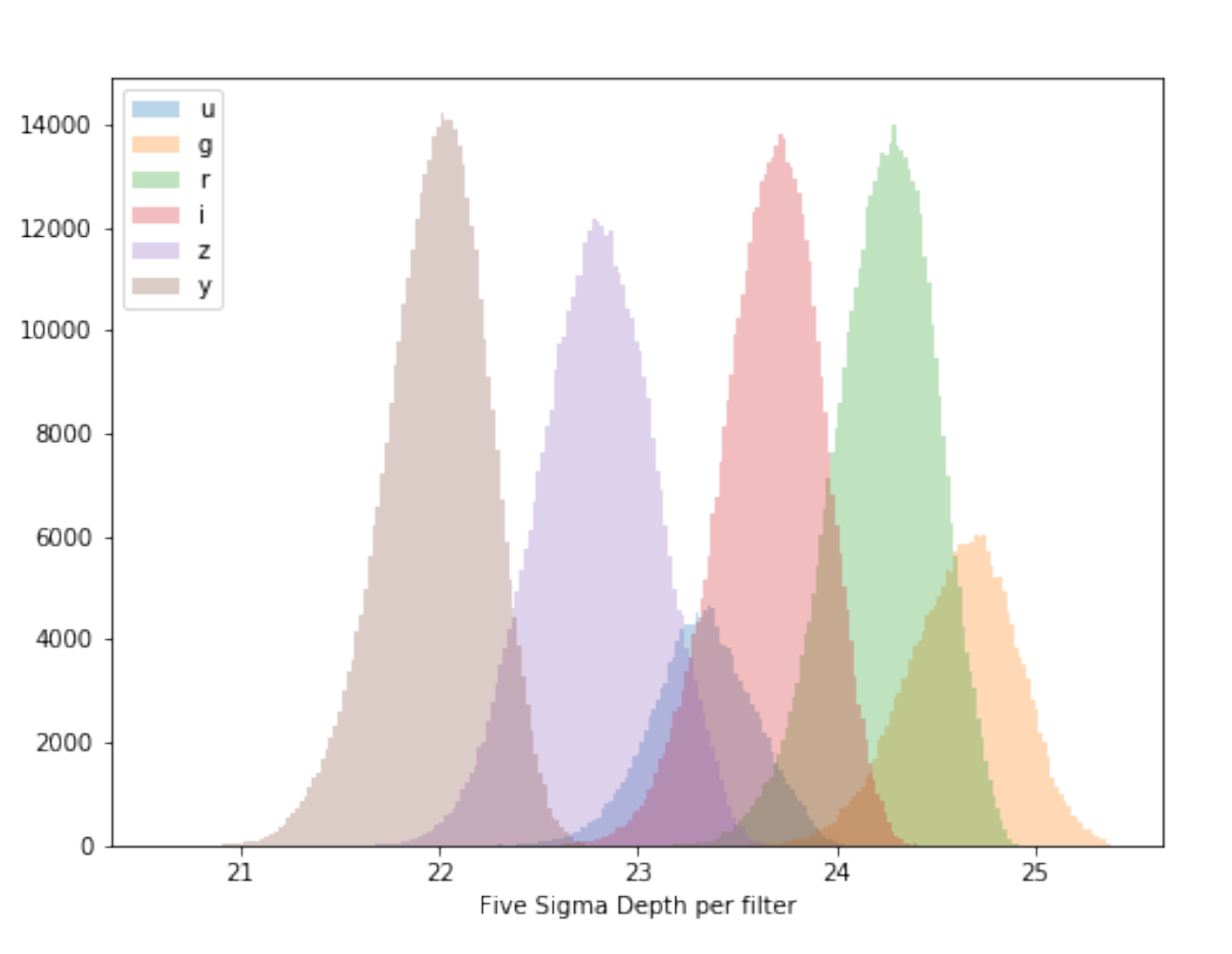}
	\vspace{-5pt}
	\caption{\label{fig:depth} Estimated distribution of five-sigma image depths in the {\it ugrizy} filters for the kraken\_2026 OpSim run. The colors of TNOs shown in Figure~\ref{fig:colossos} ($g-r=0.5-1.1$) can be used to estimate the equivalent depth of the survey in the $g$ and $r$ bands for TNOs; for $g-r\sim0.5$ (asteroids and the neutral TNOs), $g$ and $r$ achieve similar equivalent depths, but for $g-r\sim1$ (cold classical TNOs and red excited TNOs), the $r$ band images will probe 0.5 magnitudes deeper than $g$.}

	\includegraphics[width=0.7\columnwidth]{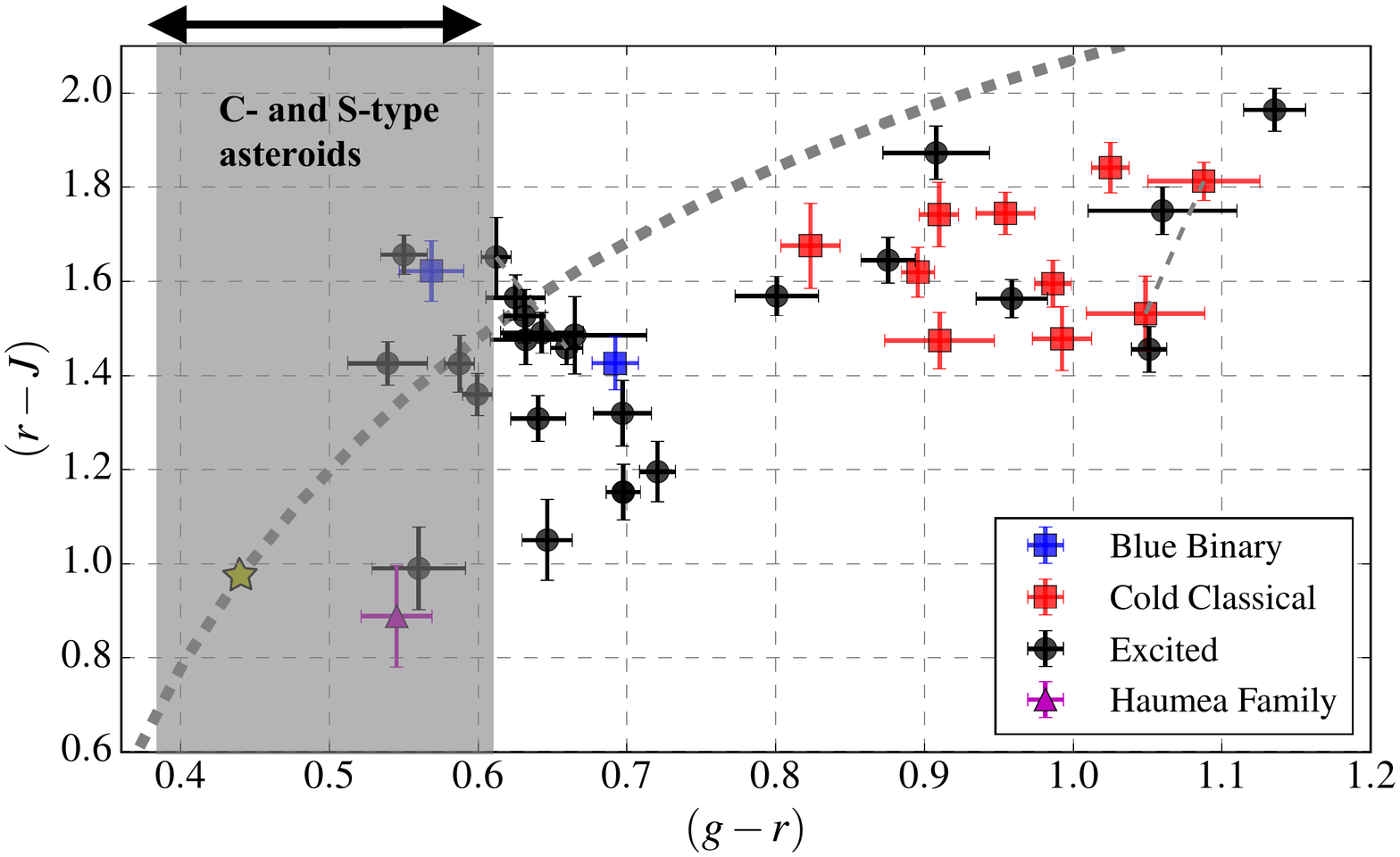}
	\vspace{-30pt}
	\caption{\label{fig:colossos} Photometric colors of a subset of transneptunian objects (TNOS)  \citep[figure modified from][]{Schwamb:2018}. The typical ($g-r$) range of C-type and S-type asteroids (the SED models currently used to represent the outer Solar System) are shown as the shaded region.}
	
   \end{center}
\end{figure}

Solar System objects are detected via reflected sunlight and are brightest in the mid-optical wavelengths. 
As outlined in the Community Observing Strategy Evaluation Paper \citep[COSEP;][]{2017arXiv170804058L}, the Wide-Fast-Deep (WFD) portion of the LSST survey will image each field in the {\it ugrizy} filters. 
Because MOPS requires a detection of Solar System objects in both images of a tracklet, the limiting magnitude for moving object detection in a given field will be set by the filter in which the Solar System objects are faintest. 
For illustration, Figure~\ref{fig:depth} shows the estimated 5-sigma image depths in each filter for the kraken\_2026 LSST operations simulation \citep[OpSim; described in][]{OpSim} run. 
The equivalent depth in each filter, and thus the optimal filter for maximizing discovery depth, will vary depending on the surface properties of different small body populations.

The intrinsic distribution of different spectral energy distributions (SEDs) for outer Solar System populations is not fully understood due to their inherent faintness. 
However several trends are apparent in their observed distribution of photometric colors.
The most prominent finding is that the surfaces of transneptunian objects (TNOs) show a bimodal distribution in $g-r$ \citep[e.g.,][]{Peixinho:2012,Fraser:2012,Peixinho:2015, Fraser:2015}, with a ``neutral" colored group ($g-r\sim0.5-0.7$) and a ``red" group ($g-r\sim0.8-1.1$) as shown in Figure~\ref{fig:colossos} adapted from \citet{Schwamb:2018}; note that both groups of TNOs are redder than solar.
These color groupings correlate with orbital dynamical properties.
Almost all of the cold classical TNOs (objects on low-inclination, low-eccentricity orbits with semimajor axes $a\sim40-45$~au) belong to the red group. The dynamically excited TNOs (objects with large eccentricities and inclinations, belonging to the hot classical, scattering, and resonant populations) span both color groups.

Discovery rates for outer Solar System populations in different LSST OpSim runs have been estimated by assuming their surfaces have the same SEDs as asteroids; these SEDs show that, in general, asteroids will be the brightest in $g$ and $r$ filters, with these two filters being roughly equivalent for discovery efficiency (i.e., simulated LSST image depths in $g$ are $\sim0.5$ magnitudes deeper than in $r$, so for asteroids with $g-r\sim0.5$, the two filters are equivalent).
However, a large portion of TNOs have surfaces that are significantly redder than the typical $g-r\sim0.5$ for asteroids (see Figure~\ref{fig:colossos}).
Thus, while the search for asteroids will achieve  similar discovery numbers in the $g$ and $r$ filters, the same will not be true for TNOs; this difference is important for future OpSim runs that include nightly pairs in mis-matched filters.

For the red group of TNOs, a comparison of Figures~\ref{fig:depth} and~\ref{fig:colossos} indicates that nightly tracklets in $g$ or a combination of $r$ and $g$ will be roughly half a magnitude shallower than nightly tracklets in $r$; due to the very steep brightness distribution of TNOs, this difference is significant.
\citet{Fraser:2014} show that the number density of TNOs as a function of magnitude can be described as $n(H)\sim10^{(0.3H)}$ in the expected limiting magnitude range for LSST. This means that the expected loss of half a magnitude in depth for the very red TNOs in $g$-band vs $r$-band would result in a $\sim75\%$ reduction in detections; this loss would impact the predominantly red cold classical population as well as the red portion of the dynamically excited population. Averaged across all TNO populations (an average $g-r=0.7$), the difference between $g$ and $r$ results in a $\sim40\%$ loss in total detections. 
This is roughly consistent with discovery rates from existing TNO surveys.\
The Outer Solar System Origins Survey \citep{Bannister:2018} surveyed in $r$-band. Near the ecliptic plane, the density of TNO detections was $\sim$2~deg$^{-2}$ at a limiting magnitude of $m_r=24.1$
\citep[based on data in Table 1 of][]{Bannister:2018}. 
The Canada France Ecliptic Plane Survey was done mostly in $g$-band, and they detected TNOs at a rate of $\sim$1~deg$^{-2}$ at a limiting magnitude of $m_g=24.4$ \citep[based on data in Table 1 of][]{Petit:2011}.

Thus we expect that nightly pairs in $g$ or in $g$ and $r$ will result in a factor of $\sim2$ fewer total detections than nightly pairs in $r$.  This potential loss when switching filters between nightly pairs of visits needs to be more accurately reflected in the Solar System metrics used for OpSim run analysis by incorporating outer Solar System SEDs.
We propose three kinds of OpSim runs for variants of the WFD survey be carried out:
\begin{itemize}
    \item an OpSim run where $r$-filter observations are always paired with other $r$-filter observations. This would represent an upper limit for detections.
    \item an OpSim run with 6 visits per month to each field in the $r$-filter, timed to meet MOPS requirements over the five months centered on opposition. This would represent an optimal search strategy in $r$ for solar system objects.
    \item an OpSim run that has a `discovery' month for each field at opposition (6 visits in the $r$-filter timed appropriately for MOPS) and at least one orbit refinement month (6 visits in the $r$-filter timed appropriately for MOPS) near quadrature. This would be a minimal search strategy in $r$ for outer solar system objects.
\end{itemize}
This will allow a robust determination of the potential loss of outer solar system detections for different filter choices in the WFD survey.

\section{Technical Description}

\subsection{High-level description}

In order to maximize the discovery of new trans-Neptunian objects, we advocate for including a sufficient number of $r$-band pairs to allow MOPS to detect and link faint, red TNOs. The optimal discovery and tracking strategy for TNOs would be to require three sets of nightly pairs in $r$ band over a 15 day period (meeting MOPS requirements) in each of the five months centered on a field's opposition. A reasonable search strategy would be to require these three sets of nightly pairs in $r$ in the opposition month and during at least one of the months near quadrature. We also advocate for including a more accurate SED for TNOs in evaluating the impact of different filter cadence options for simulations of the WFD survey using the LSST Operations Simulator \citep{OpSim}.

Adjusting the filter choice in the WFD survey to optimize TNO detections will not decrease the number of detections for other Solar System populations.
Other icy Solar System minor body populations tend to be red with respect to the Sun, but less red than the TNOs \citep{Jewitt:2015}. It is suggested that comet activity removes some of the `ultra-red matter' that coats their parent bodies; their coma colors span the range $0.45 < g-r < 0.69$ \citep[95\%-ile;][]{Solontoi2010}. Asteroids have a range of colors related to their mineralogy, but are in general less red than the comets, with colours varying from slightly blue (with respect to solar colours) to comet-like \citep{DeMeo+Carry2013}. Main-belt comet surfaces tend to be slightly bluer than solar \citep[e.g.][]{Licandro:2011}. In general, this means that the differences between sensitivity in different filters are not so extreme for inner Solar System objects, and an observing strategy that works for the detection of TNOs will also be suitable (in terms of filter choice) for other Solar System populations.

\subsection{Footprint -- pointings, regions and/or constraints}

Our proposed request for $r$ filter observations to be paired with other $r$ filter observations applies to the entire WFD footprint as well as the proposed Northern Ecliptic Spur survey, and this discussion is also briefly mentioned in the Schwamb et al. white paper. 
All classes of TNOs contain some red members, so obtaining a sufficient number of nightly $r$-band pairs for MOPS detection in all areas of the sky would maximize TNO discoveries.
The most consistently red TNOs are the population of cold classical TNOs (see Figure~\ref{fig:colossos}), which lie near the ecliptic plane.
If no $r$-band pairs were used in the WFD survey, the entire cold classical population would suffer a loss in depth of $\sim0.5$ magnitudes, resulting in significantly fewer detections. 
The dynamically excited populations are split between the red and more neutral colored populations. While their total number of detections would be decreased in the absence of $r$-band pairs, the decrease would be less dramatic because at least some of the excited TNOs will be similarly bright in either $r$ or $g$; however the reduced sample of red excited TNOs would impact science investigations into the source of the surface diversity of this population.
If $r$-band pairs are not an option for the entire WFD survey area, requiring $r$-band pairs in the area $\sim\pm10^\circ$ of the ecliptic plane would still maximize the number of cold classical TNOs discovered and provide a significant number of red excited TNO detections. 
This minimum coverage, if combined with the proposed Northern Ecliptic minisurvey for Solar System objects \citep{NES-WP}, would provide for fairly uniform detection efficiency along the entire ecliptic plane.

\clearpage
\subsection{Image quality}

No additional constraints beyond those described in Chapter 3 of the COSEP \citep{2017arXiv170804058L}.

\subsection{Individual image depth and/or sky brightness}

No additional constraints beyond those described in Chapter 3 of the COSEP \citep{2017arXiv170804058L}.

\subsection{Co-added image depth and/or total number of visits}

No additional constraints beyond those described in Chapter 3 of the COSEP \citep{2017arXiv170804058L}.

\subsection{Number of visits within a night}

At least two visits per night to a field are required to detect and identify moving Solar System objects using MOPS.

\subsection{Distribution of visits over time}
\label{sec:distvisits}

The minimum requirement on the distribution of visits over time for Solar System objects is set by MOPS detection requirements and is described in the COSEP \citep{2017arXiv170804058L}. 
The ideal distribution of visits over time for object discovery and orbit characterization is similar to that described in the Northern Ecliptic Survey white paper \citep{NES-WP}: 6 total visits to each field per month (i.e. three nightly tracklets within a 15 day period each month) in filters that optimize object detection ($r$) for five months a year (centered at opposition). However, as other science goals are likely to want visits in different filters over some of this timespan, a reasonable compromise would be to schedule a `discovery' month in the $r$-filter for each field at opposition and then one or more `orbit refinement' months in the $r$-filter separated from opposition. For outer Solar System objects, linking MOPS detections near opposition to MOPS detections near quadrature is the key to reducing orbital uncertainties, so observing a sufficient number of $r$ pairs near one or both quadrature points for MOPS to link the orbits will prevent the loss of faint TNOs that will not be detectable in observations of the field in other filters.

\subsection{Filter choice}

As discussed in Section~\ref{sec:motivation}, maximum detection efficiency for TNOs will be in the $r$ band.
For the neutral colored TNOs, comets, and asteroids, $r$ and $g$ bands are roughly equivalent given the typical depths achieved in the kraken\_2026 OpSim run (Figure~\ref{fig:depth}).
For the significant number of red TNOs, observations in $g$ will be roughly half a magnitude shallower than in $r$.
\emph{All other filters will have even less sensitivity than $g$ for TNOs.}
Observations within a single night in the WFD will not necessarily be in the same filter, however we will be constrained in nightly detection efficiency by the shallower limiting magnitude of the filter pair. 
To maximize discoveries and tracking ability with MOPS, we require that some nightly pairs be done in $r$; combinations of $r$ and $g$ are the second best option for Solar System populations.
We note that for TNOs, observations in $i$-band would potentially be of similar depth as $g$-band observations because TNOs have typical $r-i\simeq0.1-0.6$ \citep{Ofek:2012}.
However $i$-band observations would result in a shallower limiting magnitude for asteroid detections than $g$-band, so $g$ is the better choice when considering all Solar System populations.

\subsection{Exposure constraints}

No additional constraints beyond those described in Chapter 3 of the COSEP \citep{2017arXiv170804058L}.

\subsection{Other constraints}

None noted.               

\subsection{Estimated time requirement}

Our proposed changes to the WFD survey require no additional time.

\begin{table}[ht]
    \centering
    \begin{tabular}{| l | l |}
        \toprule
        Properties & Importance \hspace{.3in} \\
        \hline 
        Image quality &  2   \\
        Sky brightness & 3 \\
        Individual image depth & 1  \\
        Co-added image depth &  3 \\
        Number of exposures in a visit   & 3  \\
        Number of visits (in a night)  &  1 \\ 
        Total number of visits &  2 \\
        Time between visits (in a night) & 1 \\
        Time between visits (between nights)  & 1  \\
        Long-term gaps between visits & 2\\
        Separation between First and Final observation & 2 \\
        Filter Selection & 1 \\
        Number of Snaps in a Visit &  3 \\
        \hline
    \end{tabular}
    \caption{{\bf Constraint Rankings:} Summary of the relative importance of various survey strategy constraints, ranked from 1=very important, 2=somewhat important, 3=not important.}
        \label{tab:obs_constraints}
\end{table}

\subsection{Technical trades}

 \subsubsection{What is the effect of a trade-off between your requested survey footprint (area) and requested co-added depth or number of visits}
 
No additional trade-offs beyond those described the COSEP \citep{2017arXiv170804058L}.

 \subsubsection{If not requesting a specific timing of visits, what is the effect of a trade-off between the uniformity of observations and the frequency of observations in time? e.g. a `rolling cadence' increases the frequency of visits during a short time period at the cost of fewer visits the rest of the time, making the overall sampling less uniform.}
 
At minimum, we require the timing of the visits to be sufficient for MOPS detection and linking to prior observations. Beyond this minimum requirement, the spacing of visits in time will affect the orbit-fit quality for the detected Solar System objects. The best orbit fits will be obtained if the object is observed over a several month baseline around opposition each year. So orbit-fit quality would suffer if all of the visits to a field are, for example, in only the opposition month.

\subsubsection{What is the effect of a trade-off on the exposure time and number of visits (e.g., increasing the individual image depth but decreasing the overall number of visits)?}

No additional trade-offs beyond those described the COSEP \citep{2017arXiv170804058L}.

 \subsubsection{What is the effect of a trade-off between uniformity in number of visits and co-added depth? }
 
All of our science goals are constrained not by co-added image depth, but by the 5-sigma detection depth in individual LSST frames. This is because moving objects are found by linking the transient sources that move between visits to the field. Thus the 5-sigma limiting magnitude is of most important in the detection threshold of TNOs in the WFD survey.  There is no significant gain in trading off the uniformity in number of visits for increasing co-added depth.

\subsubsection{Is there any benefit to real-time exposure time optimization to obtain nearly constant single-visit limiting depth?}

We don't anticipate a strong impact from this and are agnostic to changes in exposure time unless the exposure time results in loss of detections for solar system objects.

\subsubsection{Are there any other potential trade-offs to consider when attempting to balance this proposal with others which may have similar but slightly different requests?\label{sec:snaps}}

Requesting same-filter nightly pairs in the $r$-filter reduces the number of chances to measure same night photometric colors for Solar System objects (and other sources) that are above the limiting magnitude in other filters that could be paired with $r$. 
Comparing photometry from different nights to obtain colors can increase the chance that ligthcurve and other time-dependent effects add noise to the color measurement.
Thus there will be a trade-off between the number of times faint TNOs are detected by MOPS (which will be maximized with $r$-band pairs no more than 90 minutes apart) and the number of same-night colors obtained for brighter Solar System objects such as asteroids that would be equally detectable in other filters.
However, our request would not affect filter combinations that do not involve $r$.
Including a more accurate SED for TNOs in the analysis of different filter strategies will allow a more accurate determination of trade-offs.

\clearpage
\section{Performance Evaluation}

The Solar System metrics described in the COSEP \citep{2017arXiv170804058L} can be used to evaluate the performance of the WFD survey for detecting TNOs with the addition of an appropriate SED for TNOs.
The simplest way to approximately incorporate a TNO SED into these metrics would be to apply a range of color corrections in the LSST filters for TNOs where they differ substantially from asteroids rather than a full SED. 
The most important color range to test is the $g-r$ color range (Figure~\ref{fig:colossos}), which shows $g-r\simeq0.5-1.1$ \citep{Schwamb:2018}. Other colors to incorporate include $r-i\simeq0.1-0.6$ \citep{Ofek:2012} and $r-z\simeq0.2-0.8$ \citep{Ofek:2012,Pike:2017}.
While we were not able to implement this prior to the white paper deadline, the authors are willing to help add TNO surface properties to the Metric Analysis Framework \citep[MAF; ][]{2014SPIE.9149E..0BJ} moving object tools in the future.

\section{Special Data Processing}

Our proposed observations have no associated special data processing requirements. The proposed change to the WFD cadence will not impact data processing. All the standard LSST data management pipelines including MOPS will be able to run on the $r$-$r$ nightly pairs.

\section{Acknowledgements}

 The authors thank the  Large Synoptic Survey Telescope (LSST) Project Science Team and the LSST Corporation for their support of LSST Solar System Science Collaboration's (SSSC) efforts. This work was supported in part by a LSST Corporation Enabling Science grant. The authors also thank the B612 Foundation, AURA, and the Simons Foundation for their support for workshops, hackathons, and sprints that lead to the development of this white paper. Elements of this work were enabled by the Solar System JupyterHub service at the University of Washington's DIRAC Institute (\url{http://dirac.astro.washington.edu}). This white paper has made use of NASA's Astrophysics Data System Bibliographic Services.

\end{document}